\documentclass[apj,iop]{emulateapj}
\usepackage{apjfonts}
\usepackage{amsmath}
\usepackage{natbib}
\bibpunct{(}{)}{;}{a}{}{,}
\DeclareMathAlphabet{\mathsc}{OT1}{cmr}{m}{sc}
\def\testbx{bx}%
\DeclareRobustCommand{\ion}[2]{%
\relax\ifmmode
\ifx\testbx\f@series
{\mathbf{#1\,\mathsc{#2}}}\else
{\mathrm{#1\,\mathsc{#2}}}\fi
\else\textup{#1\,{\mdseries\textsc{#2}}}%
\fi}

\slugcomment{Submitted to the Astrophysical Journal}

\shorttitle{Analysis of methods for detecting the proximity effect in quasar spectra}
\shortauthors{Dall'Aglio et al.}

\begin{document}

\title{Analysis of methods for detecting the proximity effect in quasar spectra}
\author{Aldo Dall'Aglio\altaffilmark{1}, Nickolay Y.\ Gnedin\altaffilmark{2,3,4}}
\altaffiltext{1}{Astrophysikalisches Institut Potsdam, An der Sternwarte 16, D-14482 Potsdam, Germany\\
  {Mail: \tt adaglio@aip.de}}
\altaffiltext{2}{Particle Astrophysics Center, Fermi National Accelerator Laboratory, Batavia, IL 60510, USA}
\altaffiltext{3}{Kavli Institute for Cosmological Physics, The University of Chicago, Chicago, IL 60637, USA}
\altaffiltext{4}{Department of Astronomy \& Astrophysics, The University of Chicago, Chicago, IL 60637 USA}

\begin{abstract}
Using numerical simulations of structure formation, we investigate several methods for determining the strength of the proximity effect in the \ion{H}{i} Ly$\alpha$ forest. We analyze three high resolution ($\sim10$~kpc) redshift snapshots ($\overline{z}=4,\ 3$ and 2.25) of a Hydro-Particle-Mesh simulation to obtain realistic absorption spectra of the \ion{H}{i} Ly$\alpha$ forest. We model the proximity effect along the simulated sight lines with a simple analytical prescription based on the assumed quasar luminosity and the intensity of the cosmic UV background. We begin our analysis investigating the intrinsic biases thought to arise in the widely adopted \emph{standard technique} of combining multiple lines of sight when searching for the proximity effect. We confirm the existence of this biases, albeit smaller than previously predicted with simple Monte Carlo simulations. We then concentrate on the analysis of the proximity effect along individual lines of sight. After determining its strength with a fiducial value of the UV background intensity, we construct the proximity effect strength distribution (PESD). We confirm that the PESD inferred from the \emph{simple averaging technique} accurately recovers the input strength of the proximity effect at all redshifts. Moreover, the PESD closely follows the behaviors found in observed samples of quasar spectra. However, the PESD obtained from our new simulated sight lines presents some differences to that of simple Monte Carlo simulations. At all redshifts, we find a smaller dispersion of the strength parameters,  the source of the corresponding smaller biases found when combining multiple lines of sight. After developing three new theoretical methods for recovering the strength of the proximity effect on individual lines of sight, we compare their accuracy to the PESD from the \emph{simple averaging technique}. All our new approaches are based on the maximization of the likelihood function, albeit invoking some modifications. The new techniques presented here, in spite of their complexity, fail to recover the input proximity effect in an un-biased way, presumably due to some (unknown) higher order correlations in the spectrum. Thus, employing complex 3D simulations, we provide strong evidence in favor of the proximity effect strength distribution obtained from the \emph{simple averaging technique}, as method of estimating the UV background intensity, free of any intrinsic biases.
\end{abstract}


\section{Introduction} 

The transition from a neutral to an ionized state of the baryonic matter in the Universe, known as the \emph{epoch of reionization}, also resulted in the appearance of the cosmic ultra-violet background radiation field (UVB). While it is still debated whether more exotic objects and processes (like mini-quasars or dark matter annihilation) had significant influence on the process of reionization \citep{haiman98,ricotti04a}, it is widely accepted that young star-forming galaxies and quasars are the primary sources of this radiation field in the post-reionization era ($z<6$). Thus, after reionization, any change in the properties of the source population is reflected in the evolution of the UV background \citep[UVB,][]{haardt96,fardal98,haardt01}. Accurate estimates of the UVB intensity at different redshifts therefore provide important constraints on the evolution of star-forming galaxies and quasars in the Universe.

The most direct probe for the UVB is the ionization state of the intergalactic medium (IGM). Mainly consisting of hydrogen and helium, the IGM becomes detectable as the light from high redshift ($z>2$) quasars travels toward us through the intergalactic space. Numerous absorption lines observed at wavelength shorter than the rest frame Ly$\alpha$ transition, known as the Ly$\alpha$ forest, arise from the small fraction of neutral matter (about 1 part in 100{,}000) in the IGM \citep{sargent80,weymann81,rauch98}. The UVB is directly responsible for keeping the IGM ionized at this level, thus encoding its intensity (and, to a lesser extent, its spectrum) in the absorption profiles of the Ly$\alpha$ forest.

Thus far only three main techniques have been employed to constrain or predict the cosmic UV background.

In the first method the hydrogen photoionization rate is measured by matching the statistical properties of the Ly$\alpha$ forest absorption obtained in numerical simulations with observations \citep{rauch97,theuns98,songaila99,mcdonald01,bolton05,giguere08b}. While this approach is extremely powerful, it has to rely on several poorly constrained quantities (such as the thermal state of the IGM), leading to large uncertainties in the evolution of the UV background \citep{giguere08b,bolton09}.

The second method is based on the determination of the overall UV luminosity density (by integrating the luminosity function) of the source populations believed to build the UV background radiation field (typically quasars and galaxies). Accounting for the filtering of the IGM with radiative transfer codes one can predict not only the evolution of the cosmic photoionization rate, but also the spectral shape of the UVB \citep{madau99,haardt01,schirber03}. The prime limitation of such a technique is due to incomplete knowledge of the galaxy luminosity function evolution at $z>3$ and the filtering power of the IGM \citep{prochaska09}, as well as poor knowledge of the escape fractions of ionizing radiation from galaxies.

Observationally, the only technique known so far to \emph{directly} infer the photoionization rate or, equivalently, the UVB intensity over some a range of wavelengths is based on the so-called \emph{proximity effect}. This effect is the manifestation of the IGM response to a systematic enhancement of UV radiation around bright quasars.

In the vicinity of a bright quasar, its UV radiation becomes several orders of magnitudes stronger than the cosmic UVB, leading to the decreased absorption blueward of the quasar Ly$\alpha$ emission line \citep{weymann81,carswell82,murdoch86}. If the quasar luminosity is known, and the relative enhancement in the UV flux near the quasar relative to the average Universe is measured from the  Ly$\alpha$ absorption spectra, the strength of the cosmic UVB can be deduced from the proximity effect \citep{carswell87,bajtlik88}. While the proximity effect has been detected for more than a decade, primarily in large samples of quasars \citep[e.g.][]{bajtlik88,lu91,giallongo96,cooke97,scott00,liske01}, recent investigations of its signature along individual lines of sight have been employed to develop a new technique for estimating the UVB intensity \citep{adaglio08,adaglio08b,adaglio09a}. 

{
In their work, Dall'Aglio et al. showed that, to a large extend, the overestimate in the UV background measurements is due to the technique itself of combining the proximity effect signal over multiple sight lines. Similar biases may affect the first attempts of constraining the average overdensity profile of typical quasars host halos \citep{rollinde05,giguere08}.
}

\begin{table}
\centering
\caption{Input parameters of the HPM simulation.}
\label{pe_theory_tab:sim}
\begin{tabular}{lclc}
\hline\hline\noalign{\smallskip}
Parameter & Value & Parameter & Value\\
\hline
$\Omega_m$        &    0.237        &  $N_p$             &    1024$^3$    \\                       
$\Omega_\Lambda$  &    0.763        &  Mesh              &    1024$^3$   \\                           
$\Omega_b$        &    0.041        &  Cell size         &    0.01$^\ddag$   \\        
$h$               &    0.735$^\dag$ &  Box size          &    10.24$^\ddag$    \\      
$\sigma_8$        &    0.742        &  $\overline{z}$    &    4.0,  3.5, 3.25,  \\
                  &                 &                    &    2.75, 2.5, 2.25  \\
\noalign{\smallskip}\hline
\end{tabular}
\begin{list}{}{}
\item[$\dag$:]  in units of $100$ km s$^{-1}$ Mpc$^{-1}$
\item[$\ddag$:]  in units of Mpc
\end{list}\end{table}

The new approach proposed by Dall'Aglio et al. is based on the analysis of the proximity effect strength distribution (PESD). Two distinct features appear in the analysis of the PESD. First, the strength distribution shows a clear peak and, second, it is significantly asymmetric. The peak of the PESD directly relates to the intensity of the UVB, whereas its asymmetry is mainly the result of low number statistics in the absorber counts near the quasar emission \citep[][hereafter Paper~II]{adaglio08b}.

This approach is nevertheless subject to a large dispersion, as it is based on the detection of the proximity effect along individual sight lines. Such a dispersion is inversely related to the change in the opacity in the Ly$\alpha$ forest, and it is further amplified by effects like overdensities or quasar variability which are poorly understood. 
It is worth mentioning that apart from studying the fluctuations of the average absorption (in terms of opacity or line number density) on single lines of sight, no other method has been developed to estimate the strength of the proximity effect. Only \citet{giguere08b} adopted a different approach using the likelihood analysis on the optical depth to characterize the mean overdensity  around high redshift quasars. However, their analysis was based not on single but on combined sight lines. 
We are therefore motivated to initiate a theoretical investigation on the methodological approach of estimating the strength of the proximity effect, primarily focusing on individual sight lines.

The plan of the paper is as follows. We begin with a description of the type of simulations employed in \S~\ref{pe_theory_txt:sims}. We then describe in detail in \S~\ref{pe_theory_txt:forest} the computation and calibration of the synthetic sight lines generated through the simulation box. \S~\ref{pe_theory_txt:proxi_eff} introduces the theoretical approach adopted to include the proximity effect on the lines of sight. We report in \S~\ref{pe_theory_txt:methods} our results for different approaches in estimating the proximity effect signature on individual objects. We then present our conclusions in \S~\ref{pe_theory_txt:conclude}.

\section{Simulations} \label{pe_theory_txt:sims}

In order to simulate moderate volumes of the Universe at high accuracy but with limited computational resources, we use the Hydro-Particle-Mesh (HPM) code developed by \citet{gnedin98}. This particular class of numerical codes differs from those following only the dark matter, in its capability of modeling both the dark matter and the baryonic components of the Universe. However, an HPM simulation is not as computationally expensive as a full hydrodynamical one.

The IGM consists of the low density cosmic gas between collapsed objects. In this low density regime there exists a tight correlation between the gas density and temperature in the form
\begin{equation}
T = T_0 (1+\delta)^{\tilde{\gamma}-1},\label{pe_theory_eq:eos}
\end{equation}
where $\delta$ is the baryonic density contrast, $T_0$ is the temperature at the mean density, which is of the order of $10^4{\rm K}$ and $\tilde{\gamma}$ ranges between 1 and 1.6. For this reason, the thermal history of the low density component of the IGM can be described with high accuracy by the evolution of the two parameters $T_0$ and $\tilde{\gamma}$. Both parameters are functions of time and are sensitive to the ionization history of the Universe. Equation~\ref{pe_theory_eq:eos}, also known as the \emph{effective equation of state}, immediately provides the thermal pressure of the gas as a function of density, thus removing the need for a full hydrodynamical solver in the code \citep{hui97,gnedin98}. 

The thermal evolution of the IGM after reionization is mainly determined by the balance between adiabatic cooling (expansion of the Universe) and photoionization heating of cosmic gas. Additional effects that influence the effective equation of state include Compton heating from X-ray sources \citep[e.g.][]{madau99b} and radiative transfer effects during \ion{He}{ii} reionization \citep[e.g.][]{maselli05}. In this work we adopt an empirical approach, and use observational constraints on the effective equation of state to ensure that the thermal state of the Ly$\alpha$ forest in our models is realistic.

The effective equation of state in the simulation was set in a piece-wise manner in three different intervals. At $z<4.5$ we used the observed evolution of $T_0$ and $\tilde{\gamma}$ \citep{ricotti00,schaye00,igm:mmrs01}. Between $z=6.5$ and $z=4.5$ we used the effective equation of state from reionization simulations of \citet{ng:gf06}; these simulations match well the observed Ly$\alpha$ opacity in the spectra of high redshift quasars discovered in the Sloan Digital Sky Survey (SDSS) and smoothly merge with the observational constraints on $T_0$ and $\tilde{\gamma}$ at $z\approx4.5$. Finally, during the reionization era ($z>6.5$) $T_0$ and $\tilde{\gamma}$ were assumed to increase linearly with the scale factor. This assumption is somewhat uncertain, but it is approximately consistent with high resolution numerical simulations of reionization \citep{ng:g04,ng:gf06} and has a negligible effect on the thermal state of the Ly$\alpha$ forest at our redshifts of interest, $2<z<4$.

For the purpose of this work we are not interested in an accurate calibration of the effective equation of state with all observational constraints, simply because the current measurements remain insufficiently precise and exhibit large scatter \citep{mcdonald01,schaye00}. For our purpose here it is sufficient that equation (\ref{pe_theory_eq:eos}) defines the underlying equation of state and that $T_0$ and $\tilde{\gamma}$ do evolve with redshift according to a specific ionization history.

\begin{figure}
\resizebox{\hsize}{!}{\includegraphics*[]{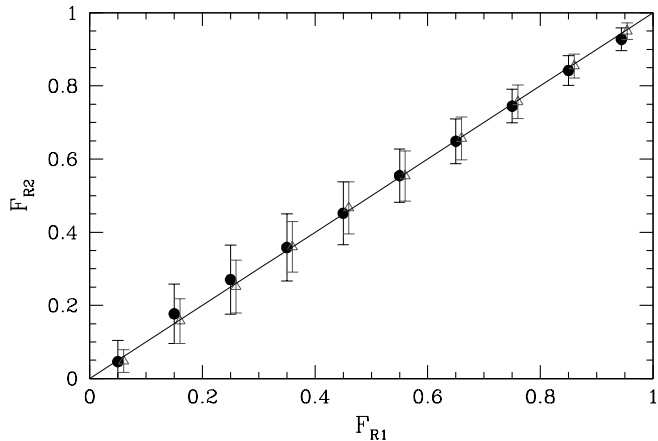}}
\caption{Convergence analysis of the mean flux in our simulated sight lines at different resolutions. The black dots represent the mean flux in a 10 comoving Mpc box and $256^3$ particles vs. $512^3$, while the gray triangles show the mean flux for $512^3$ particles vs. $1024^3$.}
\label{pe_theory_fig:resolution}
\end{figure}

Following the results of the Wilkinson Microwave Anisotropy Probe three years data \citep[WMAP3, ][]{spergel06}, Table \ref{pe_theory_tab:sim} lists the parameters adopted to generate the simulations discussed in this work. Here $\Omega_m$ is the total matter density parameter, $\Omega_\Lambda$ is the cosmological constant and $\Omega_b$ is the baryon density parameter. The Hubble constant is $h$ expressed in units of  $100$ km s$^{-1}$ Mpc$^{-1}$ and $\sigma_8$ represents the rms density fluctuation on 8 $h^{-1}$ Mpc scales at $z=0$. We fixed the box size to 10.24 $h^{-1}$ Mpc with $N_p = 1024^3$ particles on a $1024^3$ mesh. This yields a resolution element of 10 $h^{-1}$ kpc ensuring an accuracy on a few km s$^{-1}$ scale in the generation of the artificial quasar spectra (see section \ref{pe_theory_txt:spectra}). We recorded the state of the simulation of seven different redshifts denoted by $\overline{z}$.

The choice of the simulation box size is a compromise between the need to include large-scale modes and the need to fully resolve the structure in the forest. At $z=3$ our spatial resolution of 10 $h^{-1}$ comoving kpc corresponds to about 1 km/s accuracy in the velocity space. Our convergence tests, as presented in Figure \ref{pe_theory_fig:resolution}, showed that even at the resolution of 10 $h^{-1}$ comoving kpc the synthetic spectra are not fully converged. This conclusion is not in conflict with other studies, who found convergence of different statistical quantities at coarser resolutions - obviously, the requirement for a numerical simulation to converge on the synthetic spectrum on a pixel-by-pixel basis is much stricter, than, for example, to merely converge on the flux PDF. 

Our choice of 10 $h^{-1}$ comoving kpc resolution is, however, sufficient to obtain convergence to better than 0.05 in the transmitted flux at low and high flux limits, and maintain the rms precision of better than 0.1 in the transmitted flux at all flux values.

Our resolution choice results in the size of the simulation box that achieves sufficient numerical convergence on large scales for $z\lesssim3$ \citep{bb09,lidz10}, but is too short by a factor of 2 at $z\sim4$. This small loss of large scale power is unlikely to be important for our results, since the proximity effect is dominated by small scales. Never-the-less, we forewarn the reader to take our conclusions at $z\sim4$ with some caution.

\begin{figure}
\resizebox{\hsize}{!}{\includegraphics*[]{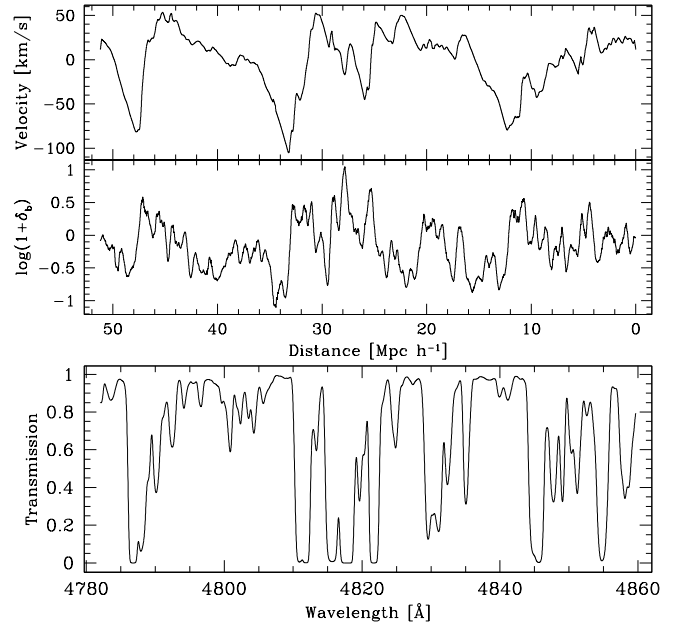}}
\caption{Example of a sight line drawn through the simulation box at redshift $z=3$. \emph{Top panel:} The peculiar velocity along the line of sight. \emph{Middle panel:} baryonic overdensity as a function of position along the line of sight. \emph{Bottom panel:} the inferred hydrogen transmitted flux as a function of wavelength.}
\label{pe_theory_fig:spec_dens}
\end{figure}

\section{The Lyman forest}\label{pe_theory_txt:forest}

\subsection{Computation of the \ion{H}{i} absorption}\label{pe_theory_txt:spectra}

The final product of an HPM simulation consists of a cosmological box (one at each $\overline{z}$), containing information about the hydrogen density contrast $\delta_\mathrm{b}$ and the relative spatial velocity ($v_x, v_y, v_z$). We use this information to compute a set of absorption spectra as follows. We draw a set of 500 randomly distributed sight lines through the box obtaining along each line of sight a spatial coordinate plus velocity and density information. In order to compute the absorption spectrum of the Ly$\alpha$ forest, we follow the methodology described in \citet{hui97}.

\begin{figure}
\resizebox{\hsize}{!}{\includegraphics*[]{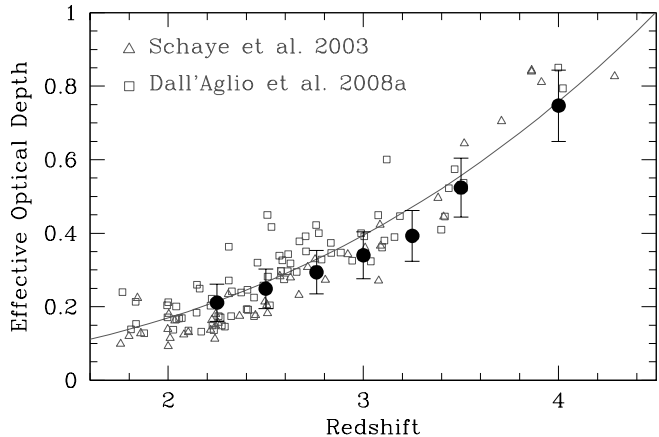}}
\caption{Effective optical depth evolution in our simulated sight lines in comparison with observations. The solid circles are the average values of $\tau_\mathrm{eff}$ in our synthetic spectra with the relative dispersions. Triangles and squares represent the measurements performed by \citet{schaye03} and \citet{adaglio08b} respectively, employing different samples of high resolution quasar spectra. The solid line represents the evolution of the effective optical depth recently estimated by \citet{adaglio08b}. }
\label{pe_theory_fig:taucomp}
\end{figure}

 The neutral fraction in the forest is determined by the balance between photoionization and recombination, and it depends both on the temperature $T$ and the intensity of the UV background $J_\ion{H}{i}$. The temperature is typically a function of the position and it is determined by the effective equation of state (eq.\ \ref{pe_theory_eq:eos}). The intensity of the UVB is, in our case, a free parameter. Due to the limited box size of the simulation, we continued each sight line for five times the length of the simulation box (i.e.\ for 50 Mpc h$^{-1}$). Because our lines of sight are oriented randomly with the respect to the box axis, that procedure does not create numerical artifacts due to the periodic boundary conditions, and we have verified that. An illustrative example of the result of our procedure is shown in Fig.~\ref{pe_theory_fig:spec_dens}. 

Finally, the absorption spectrum should match two observational constraints: (i) the evolution of the effective optical depth in the Ly$\alpha$ forest and (ii) the flux probability distribution function. To accurately calibrate our simulation, we employed the sample of 40 high resolution ($R\sim 45\ 000$), high S/N ($S/N\sim 70$) quasar spectra obtained with the UV-Visual Echelle Spectrograph (UVES), probing a redshift interval between $z\sim1.8$ and $z\sim4.6$ (Paper~II). Our simulated spectra are computed with the same spectral resolution as the observed sample, and in a similar redshift range, thus the two data sets can be directly compared.

\begin{figure*}
\resizebox{\hsize}{!}{\includegraphics*[]{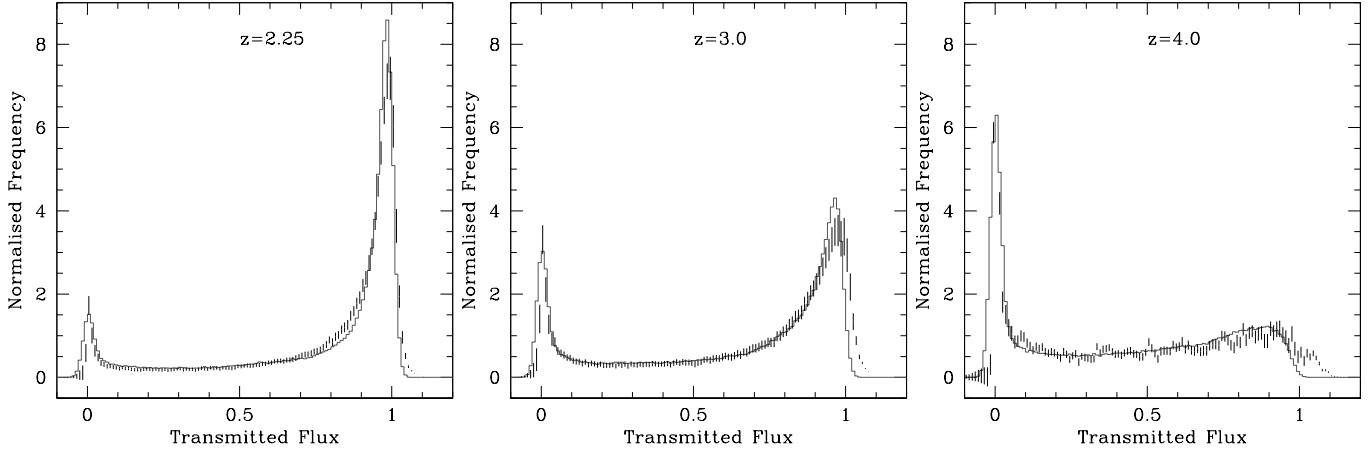}}
\caption{The average flux probability distribution (FPD) estimated from 500 simulated sight lines at three different redshifts ($z=2.25,\ 3.0$ and $4.0$, gray histogram), in comparison with the observed FPD inferred from a sample of 40 high resolution UVES/VLT quasar spectra (vertical bars). The uncertainties in the simulated FPD are negligible, while the error bars in the observed FPD account for the variance of absorption between different lines of sight and uncertainties in the continuum determination.}
\label{pe_theory_fig:pdf}
\end{figure*}

The calibration of the simulated absorption spectra has been carried out iteratively. As the main goal of this work is to test and compare different methods of estimating the proximity effect signature, we do \emph{not} attempt to match the synthetic and observed spectra exactly. Rather, we adjust the intensity of the UV background to obtain an acceptable (but not necessarily the best) match between the simulated sight lines and the observed flux probability distribution and the evolution of the effective optical depth from the UVES observations.

\subsection{The evolution of the effective optical depth}

One fundamental observed property of the Ly$\alpha$ forest is a steep decline in the hydrogen opacity towards low redshift. This behavior is reflected in the so called effective optical depth, which is defined as $\tau_\mathrm{eff}=-\ln\langle F\rangle=-\ln\langle e^{-\tau_\ion{H}{i}}\rangle$ where $F$ is the transmitted flux and the averaging $\langle\rangle$ is performed over a fixed redshift path length \citep{kim02,giguere08c}.

Because the synthetic spectra are short, $\tau_\mathrm{eff}$ does not evolve noticeably along them. Thus, we can estimate the mean $\tau_\mathrm{eff}$ and its dispersion starting from a measure of the average transmitted flux along each of the 500 simulated lines of sight and normalizing to the whole redshift interval probed by the observed spectra.

Figure~\ref{pe_theory_fig:taucomp} shows our results from the simulated sight lines.
For all snapshots at our disposal, the inferred average effective optical depth closely follows the expected enhancement at high redshift as probed by different investigations on high resolution quasar spectra \citep[][Paper~II]{schaye03}. Note that the uncertainties on the effective optical depths represent the RMS of $\tau_\mathrm{eff}$ determined on each single line of sight and not the real uncertainties of the measurements.

\subsection{The flux probability distribution}

The steep evolution of the hydrogen opacity in the Ly$\alpha$ forest described in the previous section can be detected in quasar spectra not only by measuring the average transmitted flux, but also by analyzing how the shape of the flux probability distribution (FPD) changes with redshift \citep{jenkins91}. The FPD provides a strong observational constraint, which it is important to satisfy in a realistic model of the Ly$\alpha$ forest.

We employ similar approaches to compute the FPD in the simulated and in the observed spectra. Both (the synthetic and simulated) distributions are sampled in bins of $\Delta F = 0.01$ and normalized by the bin size to maintain the condition that the FPD integrates to 1.

The observed FPD is estimated from the Ly$\alpha$ forest of those quasars intersecting a redshift slice of $\Delta z = 0.2$, centered at the redshift of the simulated snapshot ($\overline{z}$ in Tab.~\ref{pe_theory_tab:sim}). The FPD for the synthetic spectra is measured by combining the signal for all 500 lines of sight at one particular $\overline{z}$.  Additionally, we add Gaussian noise to simulated lines of sight in order to reproduce the average S/N level of the observed spectra.

Figure~\ref{pe_theory_fig:pdf} presents the comparison between the two estimates of the flux probability distribution. The agreement between the two distributions is reasonably good even if there are some indications of a departure at high redshift, in particular for the flux around unity. This lack of agreement is explained by the differences in the continuum placement of the observed and synthetic spectra. Additionally, we note that the error bars of the observed FPD are an estimate of both continuum uncertainties and Poissonian variance between different lines of sight. For the continuum uncertainties we adopted the estimates presented in Paper~II.

\subsection{The column density distribution}\label{pe_theory_txt:CDD}

The high resolution of our simulations allows us to further characterize the statistical properties of synthetic spectra by measuring the distribution of column densities. The differential distribution function of the hydrogen column densities $f(N_\ion{H}{i})$ is typically defined as the number $n$ of absorption lines per unit column density and per unit absorption path length $\Delta X$\footnote{$\Delta X=(1+z)\ \Delta z\ \left[{\Omega_\mathrm{m}(1+z)+\Omega_\mathrm{\Lambda}(1+z)^{-2}}\right]^{-1/2}$\ \citep{misawa02}} \citep{tytler87}. This distribution is typically very well represented by a single power law of the form $f(N_\ion{H}{i})\propto N_\ion{H}{i}^{-\beta}$ with $\beta$ ranging between $1.4-1.7$ \citep{hu95,kim02}.

Performing a fit of an absorption spectrum is computationally expensive, therefore we proceeded as follow: (i) we randomly select 100 simulated sight lines from our full sample of 500, (ii) for each selected line of sight we performed a Doppler profile fit using the publicly available code AUTOVP\footnote{Developed by R. Dav\'{e}: http://ursa.as.arizona.edu/$\sim$rad}, and then (iii) we visually inspected all the lines of sight in order to reject the few cases where the automatic fitting procedure fails. We repeat this procedure for sight lines drawn from the snapshot at $\overline{z}=4.0,\ 3.0$ and $2.25$ and then combine the results.

Within the range $12\lesssim \log N_\ion{H}{i}\lesssim 16$ cm$^{-2}$, the distribution accurately follows a power law with a slope of $\beta=1.64$, close to several observational results \citep{tytler87,hu95,kim02}. Our data points seem to deviate from a power law extrapolation at the low column density end. This effect, discussed in detail by \citet{hu95}, is the result of incompleteness in the sample of lines arising primarily from line blending and further amplified by noise.

\section{The proximity effect}\label{pe_theory_txt:proxi_eff}

The prime goal of this work is to continue testing different techniques for detecting the proximity effect in quasar spectra, including the Monte Carlo simulations from Paper~II. We have now a set of simulated sight lines at our disposal accurately reproducing many statistical properties of the observed Ly$\alpha$ forest. We now discuss how we introduce the proximity effect in the simulated spectra.

The optical depth of the Ly$\alpha$ forest at the observed wavelength $\lambda$ is given by
\begin{equation}
  \tau(\lambda) = \int n_\ion{H}{i} \sigma_\alpha \mathrm{d}r,
  \label{pe_theory_eq:tau_lmb}
\end{equation}
where $r$ is the physical radial coordinate along the line of sight, $n_\ion{H}{i}$ is the neutral hydrogen density at location $r$, and 
\[
  \sigma_\alpha = 4.5\times10^{-18}{\rm cm}^2 \frac{c}{b\sqrt{\pi}}e^{\displaystyle -\left(\frac{\lambda/\left(1216\AA(1+z_r)\right)-1}{b/c}\right)^2}
\]
is the cross section for the Ly$\alpha$ transition with the Doppler parameter $b$ at wavelength $\lambda$ from the gas located at position $r$, and $z_r$ is the value of cosmic redshift at that location, including the effect of the Hubble expansion and peculiar velocities. 

In the vicinity of a luminous quasar, the intensity of UV radiation produced by the quasar itself is typically up to several orders of magnitudes larger than the intensity of the UV background. This enhanced ionizing radiation acts on the neutral hydrogen which, after a period of only about 10$^4$ yr after the quasar turn-on event, reaches a new state of photoionization equilibrium. In this regime, the neutral hydrogen density of the IGM adding the quasar radiation scales as the one in its absence, by a factor of
\[
  1+\omega_r = 1 + \frac{L_H}{4\pi r^2 J_H},
\]
where $L_H$ is the hydrogen ionizing luminosity of a quasar (that we place at the origin of a line of sight $r=0$), so that the Ly$\alpha$ optical depth with the proximity effect is
\begin{equation}
  \tau(\lambda) = \int \frac{n^{\rm no PE}_\ion{H}{i}}{1+\omega_r} \sigma_\alpha \mathrm{d}r.
  \label{pe_theory_eq:tau_pe_r}
\end{equation}

In observations, however, the neutral hydrogen density $n_\ion{H}{i}$ as a function of physical distance $r$ is not directly available. Therefore, historically the proximity effect has always been approximated as
\begin{equation}
  \tau(\lambda) = \frac{1}{1+\omega_z} \int n^{\rm no PE}_\ion{H}{i} \sigma_\alpha \mathrm{d}r = \frac{\tau^{\rm no PE}}{1+\omega_z},
  \label{pe_theory_eq:tau_pe_z}
\end{equation}
where $\omega_z$ is now a function of $\lambda$ or, equivalently, cosmic redshift $z$. In other words, in all observational work the proximity effect is introduces in \emph{redshift space}, rather than in \emph{real space}. Such approximation is equivalent to ignoring peculiar velocities in the Ly$\alpha$ forest.

\begin{figure}
\resizebox{\hsize}{!}{\includegraphics*[]{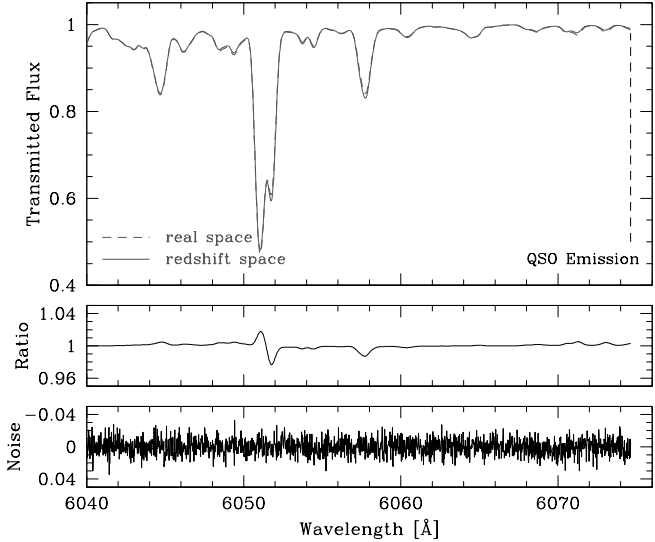}}
\caption{Example of a simulated line of sight with the proximity effect introduced in real space (dashed line) and in redshift space (solid line). The difference between the two absorption patterns are due to the peculiar velocities along the sight line, which modify the amount of ionization by changing the gas element position when the proximity effect is introduced in redshift space. The middle panel shows the ratio between the two spectra. While the influence of peculiar velocities leads to a minor discrepancy between the two spectra, this difference is drowned in the noise even in high S/N spectra, as is shown in the bottom panel (S/N=100).}\label{pe_theory_fig:pe_intro}
\end{figure}

In order to include the effect of a possible bias caused by ignoring the peculiar velocities, in the simulations we implement the proximity effect correctly, in real space, with equation (\ref{pe_theory_eq:tau_pe_r}). However, in analyzing the synthetic spectra, we follow the observational practice and assume that the proximity effect is introduced in redshift space, as in equation (\ref{pe_theory_eq:tau_pe_z}). Our measurements on the synthetic spectra, therefore, include the inevitable observational peculiar velocities bias. As an illustration, Figure~\ref{pe_theory_fig:pe_intro} shows the peculiar velocities bias along one of our lines of sight. Peculiar velocities lead to a discrepancy between the two proximity effect profiles, however this difference cannot be detected, since it is dominated by noise even in high S/N quasar spectra ($S/N \sim 100$).

Since hereafter we only consider the redshift-space proximity effect, we drop the subscript $z$ in $\omega_z$ from now on. Analytically, $\omega(z)$ can be expressed in units of the UVB photoionization rate $\Gamma_\mathrm{b}$ or in units of its intensity at the Lyman limit $J_{\nu_0}$. Following the notation of Paper~II we can write that
\begin{equation}
\omega(z)=\omega_\star \:\frac{1+z_{\mathrm{q}}}{1+z}\left(\frac{R_0}{d_{L}(z_{\mathrm{q}},z)}\right)^{2}.\label{pe_theory_eq:omega_gen}
\end{equation}
where
\begin{equation}
\omega_\star = \frac{L_{\nu_0}(z_{\mathrm{q}})} {(4 \pi R_0)^2 J_{\nu_0}}.\label{pe_theory_eq:omega_star}
\end{equation}
Here $R_0=10$~Mpc is an arbitrary distance scale introduced to make $\omega_\star$ unit-less (it also appears in Eq.~\ref{pe_theory_eq:omega_gen}). For quasars with typical Lyman limit luminosities in the range $30.5<\log(L_{\nu_0})<32.5$ and a constant UVB intensity $J_{\nu_0} = 10^{-21.51}$ in units of $\mathrm{erg}\,\mathrm{cm}^{-2}\,\mathrm{s}^{-1}\,\mathrm{Hz}^{-1}\,\mathrm{sr}^{-1}$ (Paper~II), we obtain $0.07\lesssim\omega_\star\lesssim1.5$. While our results do change quantitatively if we vary $\omega_\star$, the qualitative outcome of our analysis is independent of a particular choice of its numerical value. Therefore, by default we adopt \mbox{$\omega_\star=1$}, unless stated otherwise.

All recent results on the proximity effect have been achieved adopting the flux statistic instead of the traditional line counting approach. With the latter, it would be impossible to have a significant number of absorbers within few Mpc from the quasar to determine the proximity effect signal along single sight lines.

\citet{liske01} were the first to include the quasar proximity effect into the evolution of $\tau_\mathrm{eff}$ assuming that the optical depth scales by the same factor $1+\omega(z)$ as the neutral hydrogen density. Thus we can write that 
\begin{equation}
\tau_\mathrm{eff}=\tau_0(1+z)^{\gamma+1}(1+\omega)^{1-\beta}.\label{pe_theory_eq:taueff_omega}
\end{equation}
In the case of our simulated lines of sight, the term expressing the evolution of the effective optical depth in the Ly$\alpha$ forest, $\tau_0(1+z)^{\gamma+1}$, will be substituted by the average $\langle\tau_\mathrm{eff}(\overline{z})\rangle$ at each snapshot redshift as presented in Fig.~\ref{pe_theory_fig:taucomp}. It will be convenient to use a variable $\xi$ defined as
\begin{equation}
\xi = \frac{\tau_\mathrm{eff}}{<\tau_\mathrm{eff}(\overline{z})>} = (1+\omega)^{1-\beta},\label{pe_theory_eq:xi}
\end{equation}
where $\beta$ is the slope of the column density distribution. 

In our simulations we place the origin of all the lines of sight randomly, thus the location of the quasar (but not its emission redshift) is also random. We therefore ignore in the present analysis any effect of a biased quasar environment, i.e. overdensities around quasar locations. We do so intentionally, as our aim is to compare different methods for detecting the proximity effect along individual sight lines. Including the effect of quasar bias would add a new degree of freedom; such an effect will be addressed in an upcoming work. Furthermore, \citet{adaglio09a} presented a complete investigation of the most important biases involved in the analysis of the proximity effect such as redshift and spectrophotometry uncertainties as well as some assessment about quasar variability.

\section{Methods for estimating the proximity effect strength}\label{pe_theory_txt:methods}

\subsection{Reference approach: the combined proximity effect}\label{pe_theory_txt:std}

Among all the investigations of quasar spectra aimed at detecting of the proximity effect, two techniques have been employed so far: (i) the line counting statistic and (ii) the flux transmission statistic. Both adopt the common principle of estimating a certain quantity (number of lines or average transmission) within a regularly spaced grid in a sample of quasars. As we already showed in Paper~II the advantages of the flux transmission with respect to the line counting statistics, we will use only the flux transmission statistic as our reference technique.

For each of the simulated spectra, and given the ``input'' value $\omega_\star^{\rm IN}$, we construct the $\omega$ scale according to Eq.~\ref{pe_theory_eq:omega_gen} and then define a uniform grid in $\log \omega$ space. In each of the grid elements we determine the average flux and, thus, the effective optical depth values considering all spectra simultaneously. Finally, following Eq.~\ref{pe_theory_eq:xi}, we derive the corresponding values of $\xi$ as a function of $\omega$. The typical proximity effect signature is such that $\xi\rightarrow 0$ for $\omega\rightarrow\infty$ and it can be analytically modeled according to the formula
\begin{equation}
F(\omega)=\left(1+\frac{\omega}{a}\right)^{1-\beta},\label{pe_theory_eq:fit}
\end{equation}
where the slope of the column density distribution was fixed to $\beta=1.64$ at all redshifts according to our measurements (Sect.~\ref{pe_theory_txt:CDD}), and $a$ is a single fitting parameter. The best-fit value of $a$ can then be used to compute the ``measured'' value of the proximity effect strength $\omega_\star^{\rm OUT} \equiv a\ \omega_\star^{\rm IN}$. Ideally, this value should be close to the input value $\omega_\star^{\rm IN}$ (i.e. $a$ should be close to 1). 

This technique has been employed in the majority of the proximity effect investigation aiming at a constraint of the cosmic UV background intensity at the Lyman limit since $\omega_\star^{\rm OUT} \propto J^{-1}_{\nu_0}$. In Paper~II we first showed that this combined method is characterized by an intrinsic bias. Employing Monte Carlo simulations, we presented evidence for this bias by comparing the input and output proximity effect signal in a set of 500 synthetic spectra.

While a Monte Carlo approach may be sufficient when efficiently simulating the ``randomness'' in the properties of the absorbers, the new sight lines presented here are a significant step forward in terms of accurately reproducing the statistical properties of the Ly$\alpha$ forest. We begin our investigation comparing the results on the combined analysis of the proximity effect on both the Monte Carlo and the numerical simulated lines of sight. The Monte Carlo simulated spectra have been computed using the same procedure as in Paper~II. In all cases we employed the signal of 500 spectra including the proximity effect in the same way as described in Sect.~\ref{pe_theory_txt:proxi_eff}.

We fitted Eq.~\ref{pe_theory_eq:fit} to the values of $\xi$ determined from a combination of all sight lines. Repeating this exercise at \mbox{$\overline{z}=(2.25,\,3.0,\,4.0)$} we obtained for the Monte Carlo simulations an overestimation in $\omega_\star^{\rm OUT}$ equal to \mbox{$\Delta \log a=(0.14,\,0.1,\,0.05)$~dex}, respectively, while for the HPM simulations we obtained \mbox{$\Delta \log a=(0.1,\,0.01,\,0.01)$~dex}. This, on the one hand confirms the existence of the bias, but on the other hand shows that the Monte Carlo simulations tend to overestimate it. In particular, at $\overline{z}=3.0$ the HPM simulated sight lines predict an almost negligible overestimation. We suspect that the origin of this marginal disagreement may be primarily attribute to the procedure that generates Monte Carlo absorption spectra. The algorithm does not place a fixed number of absorption lines, instead continues to populate the spectrum with as many line as necessary to yield an evolution of $\tau_\mathrm{eff}$ consistent with a pre-fixed power law. This may then translate into a larger scatter of absorption very close to the emission redshift, thus enhancing the systematic bias when combining multiple sight lines.

\subsection{The proximity effect strength distribution}\label{pe_theory_txt:pesd}

\begin{figure}
\centering
\resizebox{\hsize}{!}{\includegraphics*[]{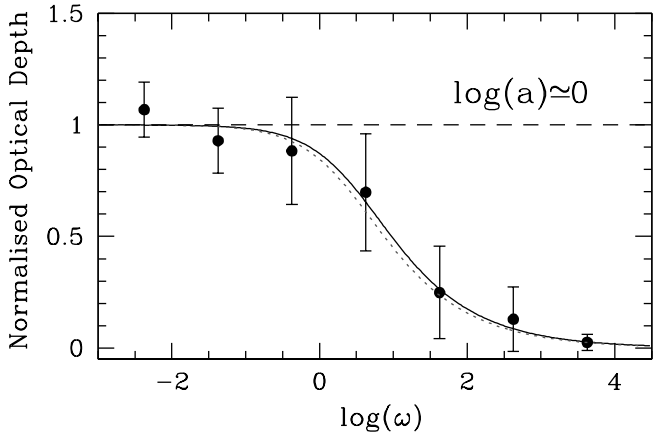}}
\caption{The proximity effect signatures in one simulated line of sight. The data points show the normalized effective optical depth $\xi$ versus $\omega$, binned in steps of $\Delta \log \omega = 1$. The dotted line represents the reference model used to introduce the proximity effect in the synthetic spectrum. The solid line shows the best fit model as described in Sect.~\ref{pe_theory_txt:std}}
\label{pe_theory_fig:pe_plot_ex}
\end{figure}

A correct understanding of the biases involved in the combined analysis of the proximity effect is essential to accurately determine the cosmic UV background intensity. We proposed in Paper~II a new technique of measuring the UVB intensity, unaffected by the biases described in the previous section. This approach is based on the determination of the proximity effect along individual lines of sight in a quasar sample. Always adopting Monte Carlo simulated lines of sight at different redshifts, they fitted equation (\ref{pe_theory_eq:fit}) to individual spectra and showed that
\begin{enumerate}
\item{the distribution of $\log \omega_\star^{\rm OUT}/\omega_\star^{\rm IN}\equiv \log a$ is skewed}
\item{the skewness increases with decreasing redshift}
\item{this asymmetry is the main contributor to the overestimation of the UVB found in the literature}
\item{the peak of this distribution  is an unbiased estimate of the UV background intensity}
\end{enumerate}

The skewness of the proximity effect strength distribution (PESD) originates from the definition of the uniform grid in $\log \omega$ space. In other words, as a constant $\log \omega$ range progressively probes smaller redshift intervals approaching the quasar, the absorbers tend to no longer be Gaussian distributed. This effect is further enhanced at lower emission redshifts since the line number density decreases. Therefore, the distribution not only becomes broader, but also more skewed.

\begin{figure*}
\centering
\includegraphics*[width=12cm]{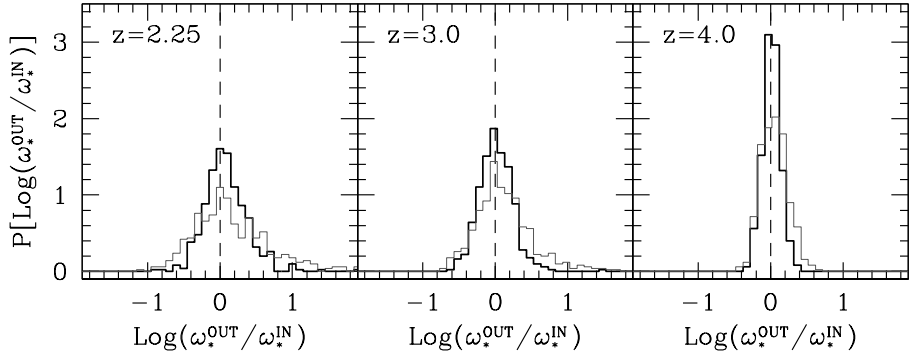}
\caption{The proximity effect strength distribution (PESD) in three different sets of 500 sight lines drawn from our HPM simulation boxes at redshift $\overline{z}=2.25,\ 3.0$ and $4.0$ (thick histogram). The thin histogram represents the PESD obtained from a sample of 500 Monte Carlo simulated lines of sight. For both types of simulations we determined the proximity effect strength adopting the best-fit $\log a$ value of Eq.~\ref{pe_theory_eq:fit}. The vertical dashed line marks the reference model used for creating synthetic Ly$\alpha$ forest spectra.}
\label{pe_theory_fig:std_avg}
\end{figure*}

To check how accurately we can recover the input value $\omega_\star^{\rm IN}$, we fit Eq.~\ref{pe_theory_eq:fit} to all 500 lines of sight at three different redshifts ($z_\mathrm{q}=2.25,\ 3$ and $4$). That gives us an estimate of the proximity effect strength $\omega_\star^{\rm OUT}$ along each sight line. Figure~\ref{pe_theory_fig:pe_plot_ex} illustrates a typical example of the proximity effect signature along one sight line in our HPM simulations. All lines of sight can then be combined to form the proximity effect strength distribution. Figure~\ref{pe_theory_fig:std_avg} presents our results.

We confirm with our simulations the results reported in Paper~II: the PESD sharply peaks at the input model ($\log \omega_\star^{\rm OUT}/\omega_\star^{\rm IN}=0$) and becomes broader towards lower redshift. Furthermore, the skewness in the PESD increases towards low redshift. However, our results on the PESD inferred from the HPM simulation quantitatively differ from the Monte Carlo simulations. While the peaks of the distributions match, the rms are significantly smaller for the HPM-based sight lines. We obtained at redshifts \mbox{$\overline{z}=(2.25,\,3.0,\,4.0)$} a dispersion of strength parameter equal to \mbox{$\sigma \log a=(0.3,\,0.23,\,0.1)$~dex} for the HPM simulations, while  in the Monte Carlo one we estimated \mbox{$\sigma \log a=(0.65,\,0.5,\,0.2)$~dex}. The larger dispersion in the latter results in the stronger bias in the combined proximity effect analysis reported in the previous section. 

To precisely estimate the uncertainties related to the modal value of the PESD we adopted a bootstrap technique. Starting from a distribution of $N_i$ values of $\log a$, where $N_i$ represents the total number of $\log a$ estimates, we randomly duplicated $N_i/e$ strength parameters and estimated the modal value of the new PESD. We repeated this process 500 times for each redshift snapshot (as well as in the following), obtaining the mean and the sigma values of PESD modes.

Measuring the proximity effect signal along individual lines of sight, and thus determining the PESD, allows unbiased estimates of the cosmic UV background. However, this method is still based on a simple averaging process of the absorption in the Ly$\alpha$ forest. In other words, the advantage of dealing with very high quality data is not fully explored. Hereafter, we will refer to the PESD estimated from the normalized optical depth on individual lines of sight as the \emph{simple averaging technique}.

\begin{figure}
\resizebox{\hsize}{!}{\includegraphics*[]{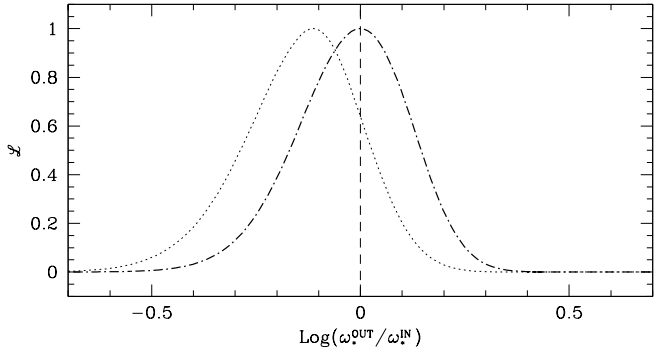}}
\caption{An example of the likelihood function $\mathcal{L}$ estimated from two different sight lines in our simulations. While in one sight line (dotted dashed profile) the likelihood is maximized at the input model, the second sight line  (dotted profile) has the most likely value of $\omega_\star^{\rm OUT}$ significantly below $\omega_\star^{\rm IN}$.}
\label{pe_theory_fig:LL_ex}
\end{figure}
\begin{figure*}
\centering
\includegraphics*[width=12cm]{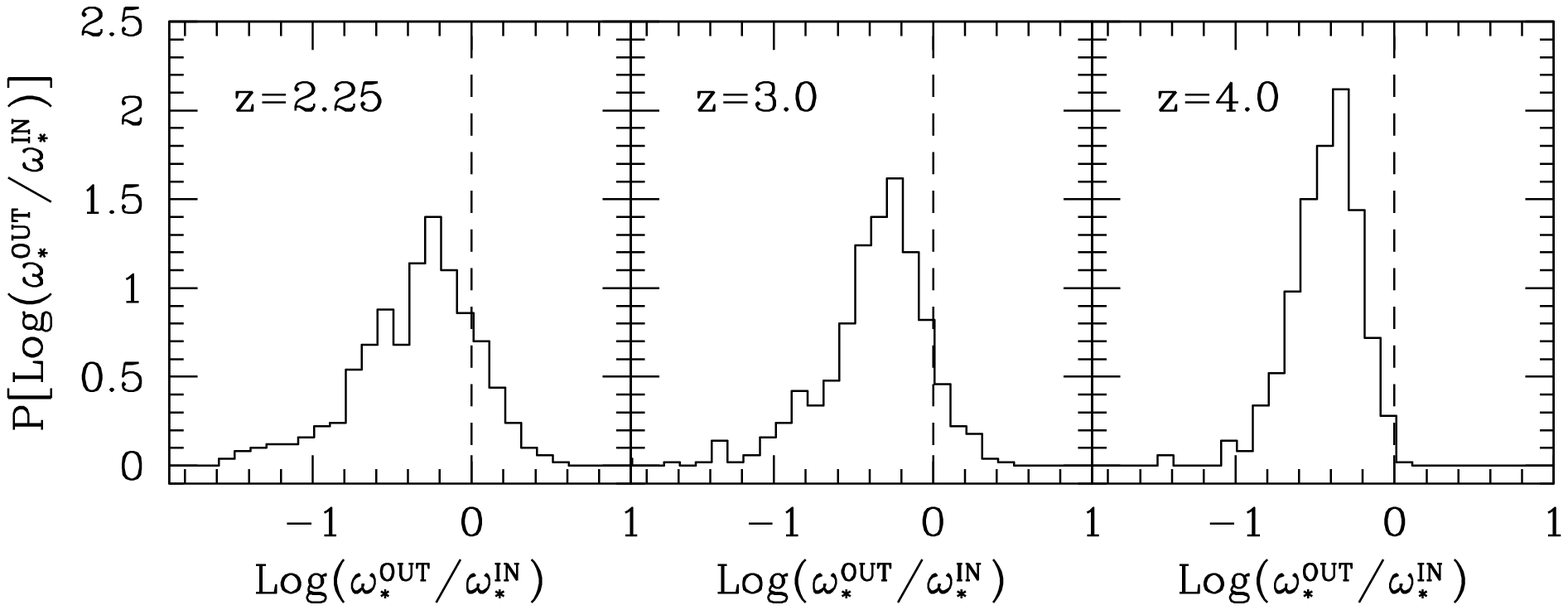}
\caption{The proximity effect strength distribution in three different sets of 500 sight lines at redshift $\overline{z}=2.25$, $3.0$, and $4.0$. The PESD has been constructed adopting the likelihood technique described in Sect.~\ref{pe_theory_txt:LL_v0} to estimate the strength of the proximity effect. The extent of the biases, represented by the shift of the mode in the PESD with respect to the dashed vertical line, remains constant with redshift.}
\label{pe_theory_fig:LL_v0}
\end{figure*}

\subsection{The maximum-likelihood approach}\label{pe_theory_txt:LL_v0}

The importance of a precise determination of the UVB intensity at different epochs motivates us to further develop and test \emph{new} methods of determining the proximity effect strength.

A widely used, extremely flexible approach for recovering input parameters is the maximization of the likelihood function (LF). It expresses the probability of a set of parameters in a statistical model describing certain data. In our case, we can write this function as the probability that our spectrum has been modified by the quasar radiation of a given strength $\omega_\star$.

Generally the likelihood function is defined as
\begin{equation}\label{pe_theory_eq:eqL0}
\mathcal{L} =\prod_{i=1}^N P(F_i|C)
\end{equation}
where the product is calculated over $N$ data points, and $P(F_i|C)$ is the probability of occurrence of the measurement $F_i$ given the set of parameters $C$. Here, all data points $F_i$ are flux values in the observed or synthetic Ly$\alpha$ spectrum. 

The prime limitation of Eq.\ \ref{pe_theory_eq:eqL0} is that the product operator must be applied to uncorrelated data points. However, neighboring pixels in the Ly$\alpha$ absorption spectrum are strongly correlated due to both physical correlations of cosmic large-scale structures and thermal and instrumental broadening of absorption lines. Eq.\ \ref{pe_theory_eq:eqL0} can be generalized for the case of correlated data, but that would require knowing an $N$-point correlation function for the flux, which is possible to estimate in any reasonable way neither from the observational data nor from the simulations.

Therefore, as a first attempt, we adopt a simplified approach and re-bin the spectra over at least 40 km s$^{-1}$ (the average width of an absorber in the velocity space) in order to significantly reduce the correlation in the Ly$\alpha$ forest without losing too much resolution. In the following sections we discuss more sophisticated methods of accounting for the correlations between the data points. 

Let us now compare the same sight line with and without the signature of the proximity effect. These two spectra have the same original hydrogen distribution along the line of sight, thus the two hydrogen densities, or the two optical depths, are proportional by a factor of $1+\omega(z)$ as discussed in Sect.~\ref{pe_theory_txt:proxi_eff}. Our aim is to express the observed flux probability density $P(F)$ in Eq.~\ref{pe_theory_eq:eqL0} as a function of the strength of the proximity effect and the flux probability unaffected by the quasar radiation.

The FPD affected ($P_m$) and unaffected ($P_\infty$ as presented in Fig.~\ref{pe_theory_fig:pdf}) by the quasar radiation are related by
\begin{equation}\label{pe_theory_eq:eqL1}
P_m(F_m)\mathrm{d}F_m = P_\infty(F_\infty)\mathrm{d}F_\infty.
\end{equation}
Knowing that
\begin{equation}\label{pe_theory_eq:eqL2}
F_m = \mathrm{e}^{-\tau_m} = \mathrm{exp}\left(- \frac{\tau_\infty }{1+\omega}\right)=F_\infty^{1+\omega},
\end{equation}
we can write  
\begin{equation}\label{pe_theory_eq:eqL3}
P_m(F_m) = P_\infty\left(F_m^{1+\omega}\right)(1+\omega)F_m^{\omega}.
\end{equation}
The Likelihood function in Eq.~\ref{pe_theory_eq:eqL0} can be generalized in the presence of instrumental noise to
\begin{equation}\label{pe_theory_eq:eqL4}
\mathcal{L} =\prod_i\int_0^1\frac{\mathrm{exp}\left[-(F_{i,m}- F')^2/2\sigma_i^2\right]}{\sigma_i \sqrt{2\pi}} P_m(F')\mathrm{d}F'
\end{equation}
where the additional exponential term describes the Gaussian noise with the proper normalization. Inserting Eq.~\ref{pe_theory_eq:eqL3} into  Eq.~\ref{pe_theory_eq:eqL4} we obtain
\begin{eqnarray}\label{pe_theory_eq:eqL5}
\hspace{1.2cm}\nonumber\mathcal{L}&=& \prod_i \int_0^1 \frac{\mathrm{exp}\left[  -(F_{i,m} - F')^2 / 2\sigma_i^2\right]}{\sigma_i \sqrt{2\pi}}\times\\
&\times& P_\infty\left(F'^{1+\omega}\right)(1+\omega)F'^{\omega}\mathrm{d}F'.
\end{eqnarray}
This function has only one free parameter, the strength of the proximity effect $\omega_\star^{\rm OUT}$. For an infinitely high signal to noise ratio, the Gaussian becomes a delta function and the expression under the integral reduces to the flux probability distribution given by Eq.~\ref{pe_theory_eq:eqL3}.

\begin{figure}
\resizebox{\hsize}{!}{\includegraphics*[]{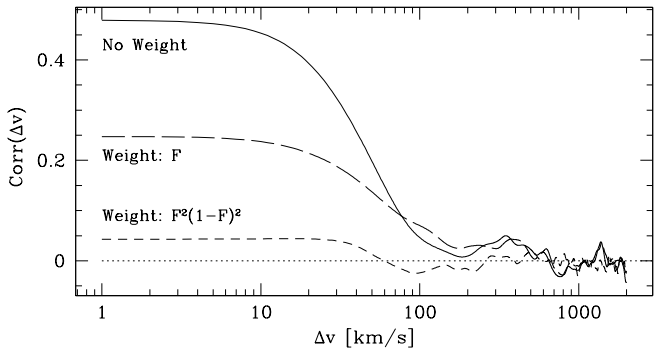}}
\caption{The flux auto-correlation function in one Ly$\alpha$ forest spectrum (solid line) in comparison with two different non-trivial weighting schemes (long and short dashed lines). The introduction of a weighting scheme in the computation of $\mathrm{Corr}(\Delta v)$ significantly reduces the auto-correlation of the transmitted flux.}
\label{pe_theory_fig:corr_los}
\end{figure}

\begin{figure*}
\centering
\includegraphics*[width=12cm]{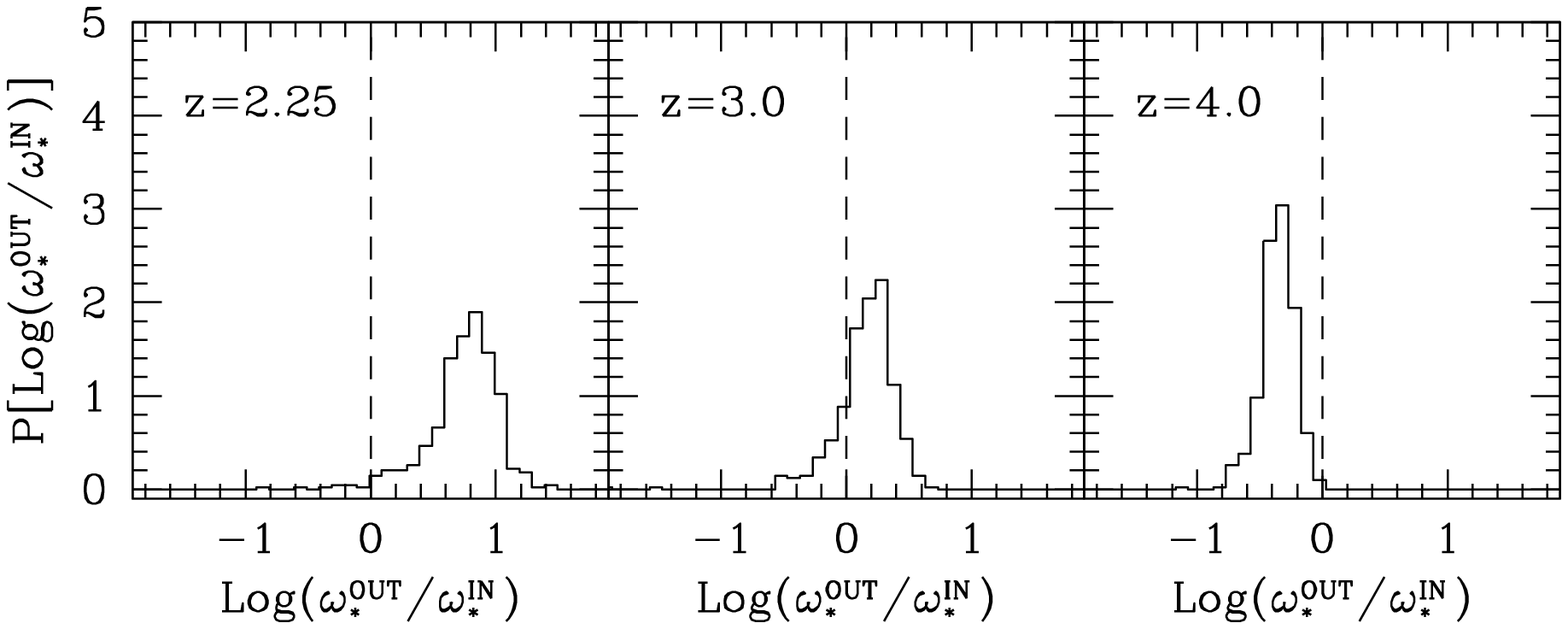}
\caption{The proximity effect strength distribution in three different sets of 500 sight lines at redshift $\overline{z}=2.25$, $3.0$, and $4.0$. The PESD has been constructed with a modified likelihood approach, which used a weighted FPD to account for the intrinsic correlations in the Ly$\alpha$ forest. The PESD remains biased with respect to the reference model and the amount of bias is now strongly redshift dependent.}
\label{pe_theory_fig:wpdf_nosampl}
\end{figure*}

The methodological approach is then straightforward: (i) we introduce the proximity effect on the hydrogen neutral fraction along all the lines of sight at our disposal, (ii) we compute the likelihood function for a set of $\omega_\star^{\rm OUT}$ values within the range $-1<\log \omega_\star^{\rm OUT}<1$ and finally (iii) we search for the particular value of $\omega_\star^{\rm OUT}$ that maximizes Eq.~\ref{pe_theory_eq:eqL5}. For illustration purposes we present in Fig.~\ref{pe_theory_fig:LL_ex} the likelihood function for two different lines of sight where the two maxima are located in different positions with respect to the input model. After repeating this procedure over all spectra we construct the PESDs presented in Fig.~\ref{pe_theory_fig:LL_v0}. 

At all redshifts the inferred PESD has a clear maximum, however this maximum does not coincide with the input model, moreover the modal values of all PESDs are biased towards smaller $\omega_\star^{\rm OUT}$. Contrary to the outcome of the \emph{simple averaging technique}, this approach fails to recover the input model and also the inferred PESD is clearly broader. Several factors may cause this bias.

(i) Our spectra might still be significantly affected by intrinsic (as opposed to thermal or instrumental broadening) correlations in the Ly$\alpha$ forest. Even when we re-bin the spectrum to significantly reduce the correlations between nearby pixels, the intrinsic correlations between close absorbers largely remain. For this reason we recomputed the PESD after re-binning the spectrum over several tens of km s$^{-1}$, up to 100 km s$^{-1}$. Unfortunately this had no effect neither on the modal value of the PESD nor on its shape, demonstrating that it is intrinsic correlations between absorbers and not thermal broadening that is responsible for the biased result of Fig.~\ref{pe_theory_fig:LL_v0}.

(ii) The flux probability distribution might change significantly along different lines of sight, thus concealing some uncontrolled systematic effect when assuming as common FPD the average over all sight lines. Therefore we have repeated our computation adopting the FPD estimated from the same line of sight without the influence of the quasar. Such a procedure is, of course, not feasible for real observations, but, nevertheless, it does not solve the problem of a biased PESD. We have also tried to analytically fit the average or single FPD with different fits \citep{miralda-escude00,becker07}, but such procedure does not lead to a smaller bias.

The latter approach is similar to that of \citet{giguere08}. They adopted the likelihood method, assuming a log-normal distribution, to derive the average halo mass hosting a quasar. The main difference is that we are applying this likelihood method on single sight lines to characterize more accurately the fluctuations of the density field near the quasars.

We conclude that the reason for the bias in the maximum likelihood analysis is caused by an intrinsic correlation in the Ly$\alpha$ forest not being accurately accounted for by our simple re-binning procedure. In the following, we attempt to solve this problem by estimating the correlation function in our simulated spectra.

\subsection{The correlation function}\label{pe_theory_txt:correlation}

We showed in the previous section that clustering of Ly$\alpha$ absorbers gives rise to correlations in the observed transmitted flux large enough to heavily bias the results of a maximum likelihood analysis, even after re-binning the simulated spectrum. We now focus on measuring how large these correlations are by means of the \emph{correlation function}.

Given a point in redshift with transmitted flux $F$, the correlation function describes the probability of finding another point, with the same $F$, within a given redshift interval. More precisely, if we express the interval with a velocity shift $\Delta v$ we can write that
\begin{equation}
\mathrm{Corr}(\Delta v) = \left\langle\delta F(v) \delta F(v+\Delta v)\right\rangle/\overline{F}^2, \label{pe_theory_eq:corr}
\end{equation}
where $\overline{F}$ represents the mean flux and $\delta F(v)\equiv F(v)-\overline{F}$. The numerical value of $\mathrm{Corr}(\Delta v)$ is obtained by directly averaging individual pixels over the spectrum, separated by a given velocity $\Delta v$.

To illustrate the properties of the correlation function, we computed $\mathrm{Corr}(\Delta v)$ for one sight line and show the result in Fig.~\ref{pe_theory_fig:corr_los}. The amplitude of the correlation increases significantly for separations smaller than about $\Delta v\lesssim 100$ km s$^{-1}$, while it fluctuates around zero for separations larger than a few hundreds of km s$^{-1}$. This directly shows that individual pixels in the Ly$\alpha$ forest are not independent from an other, but strongly correlated. Such a correlation, which also changes from one sight line to another, does not vanish after a simple re-binning of the spectrum.

Motivated by the lack of success in our previous method, we explore a different approach to remove the signature of correlated pixels. We introduce a weighting scheme in the definition of the correlation function designed to give negligible weight to correlated pixels in the  Ly$\alpha$ forest. Adopting this weight to estimate a new flux probability distribution, we would immediately remove the imprints of the correlation.

If we introduce such a weighting scheme, Eq.~\ref{pe_theory_eq:corr} becomes
\begin{equation}
\mathrm{Corr}(\Delta v,w) = \frac{ \left\langle\delta F(v)\delta F(v+\Delta v)\right\rangle_w }{ \left\langle F(v)\,\right\rangle_w\left\langle F(v+\Delta v)\right\rangle_w}.\label{pe_theory_eq:corr2}
\end{equation}
For this purpose, we explored two types of weighting functions: $w_1(F) = F$, which removes the correlations for strong absorbers and, $w_2(F) = F^2(1-F)^2$, which accounts for the correlation of strong and weak absorbers. With our new definition of the correlation function, we recomputed $\mathrm{Corr}(\Delta v,w)$ for the same sight line as before and place our results into context in Fig.~\ref{pe_theory_fig:corr_los}. While already the first weight significantly reduces the correlations, the second one removes the intrinsic correlations of the the Ly$\alpha$ flux almost completely.

We then adopted $w_2(F)$ to recompute the FPD which will now have a different shape with respect to that of Fig.~\ref{pe_theory_fig:pdf}, and will show one pronounced peak for $0<F<1$. This new weighted probability distribution is used to infer the likelihood function following the same procedure as in Sect.~\ref{pe_theory_txt:LL_v0}. From the most likely values of the proximity effect strength, we reconstructed the PESDs which are now presented in Fig.~\ref{pe_theory_fig:wpdf_nosampl}. With this new approach, all the inferred distributions not only present a significant bias with respect to the input model, but this bias additionally changes from an underestimation to an overestimation as the snapshot redshift decreases. In spite of the complexity of this new method, there are still uncontrolled systematics in the analysis of the proximity effect which are not correctly accounted for, even introducing a weighting scheme.

\begin{figure*}
\centering
\includegraphics*[width=12cm]{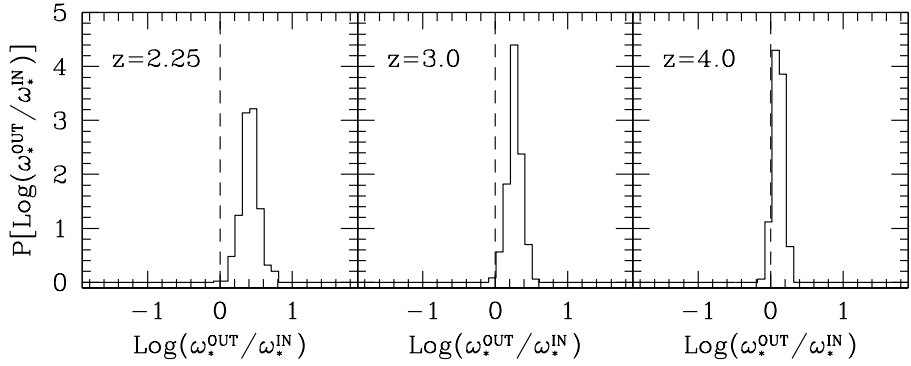}
\caption{The proximity effect strength distribution in three different sets of 500 sight lines at redshift $\overline{z}=2.25$, $3.0$, and $4.0$. The PESD has been constructed employing a modified likelihood approach which adopted the unweighted FPD but samples the spectrum over 2000 km s$^{-1}$ to account for the absorption correlations in the Ly$\alpha$ forest. The PESD remains biased with respect to the reference model and the amount of bias is now strongly dependent on redshift.}
\label{pe_theory_fig:LL_sampl}
\end{figure*}
\subsection{Sampling the Ly$\alpha$ forest for the likelihood}

None of the techniques presented so far performs better than the simple averaging technique in recovering the signature of the proximity effect. Our next attempt to overcome the imprints of the mentioned correlations is based on  the computation of a different likelihood function. 

Until now we have proceeded with the computation of $\mathcal{L}$ following Eq.~\ref{pe_theory_eq:eqL5}, where the product is performed considering all the flux pixels in the spectrum. Due to the absorption correlations and our difficulties in removing their signature, we now try to apply a selection of the pixels from which the product will be estimated. If we consider a set of $i$ flux pixels separated by a few thousands of km s$^{-1}$ ($\Delta v$), these points will be uncorrelated according to Fig.~\ref{pe_theory_fig:corr_los}. From this set of flux values we can estimate one likelihood function before considering to the next set of $i$ flux pixels.

Depending on the resolution of our synthetic spectra ($\mathrm{d} v$), we will have a set of $j$ likelihood functions where the exact number is defined as $j = \Delta v / \mathrm{d} v$. Each  likelihood function will then be maximized and yield one value of $\omega_{\star, j}^{\rm OUT}$. 

The distribution of $\omega_{\star, j}^{\rm OUT}$ depends on how many points contribute to the particular set and behaves as follows: increasing the pixel separation $\Delta v$, the number of pixels from which the likelihood is estimated decreases, thus yielding a broader distribution of $\omega_{\star, j}^{\rm OUT}$. Equivalently, if the pixel separation is too small, the influence of the correlation between pixels increases, resulting is a biased result. We fix our separation to $\Delta v=2000$ km s$^{-1}$ and adopt the mean $\omega_{\star, j}^{\rm OUT}$ as a proxy for the most likely indicator of the  proximity effect strength.

Figure~\ref{pe_theory_fig:LL_sampl} presents the results showing the PESDs at different redshifts. In our highest redshift snapshot the modal value of the PESD is, given the uncertainties, extremely close to the input model (offset by $1.4\sigma$) with a dispersion in the strength parameter significantly smaller than that of the simple averaging technique. However, towards lower redshift, a significant bias in the modal values appears again. We conclude that this approach is not superior to the simple averaging technique.

Even if we could find two new pixel separations $\Delta v$ at redshifts $\overline{z}=2.25$ and $3.0$ which yield no biases in the recovery of the reference model, we are aware of the drawbacks that such approach would have on real data. The biases that are in this method due to a \emph{wrong} choice of the sampling size are extremely difficult to control because in real spectra we typically lack of the spectral informations without the quasar impact. At best, we therefore could only guess the \emph{appropriate} sampling size via numerical simulations, without being able to test the accuracy.

\begin{figure}
\resizebox{\hsize}{!}{\includegraphics*[]{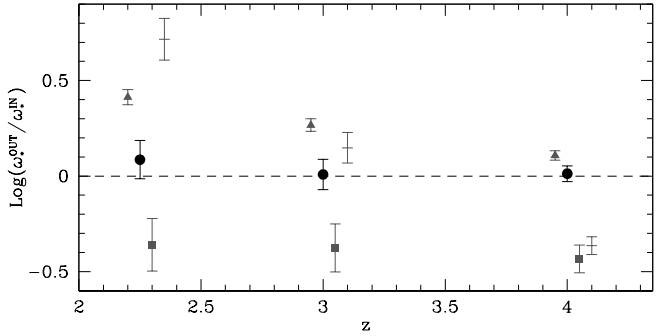}}
\caption{Comparison of the best estimate of the proximity effect strength obtained with the different methods presented in this work. The black circles show the outcome of the simple averaging technique. The squares, the crosses and the triangles depict the simple likelihood, the likelihood with a weighted FPD, and the sampled likelihood, respectively. A small redshift shift has been applied to the data points to make them more easily recognizable. The modal values and associated errors have been estimated with a bootstrap technique as described in Sect.~\ref{pe_theory_txt:methods}.}
\label{pe_theory_fig:overview}
\end{figure}

We summaries the results of all the techniques presented in this work in Fig.~\ref{pe_theory_fig:overview}. None of the more maximum likelihood methods are capable of yielding tighter and unbiased constraints on the proximity effect than the simple averaging technique.

\section{Conclusions}\label{pe_theory_txt:conclude}

We have analyzed a set of high-resolution, three-dimensional numerical simulations with a Hydro-Particle-Mesh code. We evolved the particle distribution in the simulated box until a redshift of $\overline{z}=2.25$, and recorded seven snapshots within the range $2.25<\overline{z}<4$. For each snapshot we have drawn 500 randomly distributed sight lines through the simulated box, obtaining simulated spectra of the Ly$\alpha$ forest. 

A sample of 40 high-resolution, high-S/N quasar spectra, with emission redshifts within the range $2.1<\overline{z}<4.7$, has been used to calibrate the simulated spectra. We have computed from the simulated sight lines (i) the evolution of the effective optical depth, (ii) the flux probability distribution function, and (iii) the column density distribution at different redshifts. While the computation of the synthetic line of sight depends on several free parameters, we have tuned them to be consistent (within the measured uncertainties) with the observational data in all three measurements.

Our study is focused on developing and testing new techniques of recovering the strength of the proximity effect along individual sight lines. Our analysis has begun with a comparison between the widely adopted \emph{combined analysis} of the proximity effect signal over multiple lines of sight, with the recently developed technique of estimating its strength on individual quasar spectra. We refer to this method the \emph{simple averaging technique}. As the strength distribution is supposed to be asymmetric, biases are expected to arise when determining the combined proximity effect signal.

We have confirmed, with a realistic set of synthetic lines of sight drawn from our numerical simulation, the existence of this biases, albeit with a different intensity as predicted with Monte Carlo simulations. We have concluded that the smaller bias is caused by the smaller scatter of the strength parameter. We have confirmed that the modal value, or peak of the proximity effect strength distribution (PESD), yields an unbiased estimate of the input parameters used to compute the proximity effect. Moreover, we have detected the expected broadening in the shape of the PESD towards low redshift, as predicted in Paper~II using Monte Carlo simulations.

In principle, the simple averaging technique, by combining observed pixels together, loses information. In order to avoid this loss, we have investigated several incarnations of a maximum likelihood approach. The first incarnation was a standard implementation of the likelihood function. Due to intrinsic (as opposed to thermal and instrumental broadening) auto-correlation of the transmitted flux along a single line of sight, this technique was subject of systematic bias at all redshifts and for all models of the flux probability distribution.

In the second method we have used a weighting scheme, designed to reduce the intrinsic auto-correlation in the absorption spectrum. While this weighting scheme was able to substantially reduce the two-point autocorrelation functions of the flux, the resultant PESDs were significantly biased. This failure of the weighting scheme indicates that it is not a two-point, but some (currently unknown) higher order correlation function(s) that are primarily responsible for the bias in the maximum likelihood estimate of the proximity effect.

In an attempt to reduce the bias, we have adopted a sampling approach of widely separated flux points in the spectrum to design a more complex likelihood function. While this approach yielded a substantially more accurate estimate of the best-fit value, the value itself remained biased. That bias is comparable to the statistical uncertainty of the measurement at redshift $\overline{z}=4.0$, but becomes progressively larger towards lower redshifts.

Thus, the newly introduced  \emph{simple averaging technique}, despite of the perceived loss of information during the averaging procedure, is the only method of estimating the proximity effect signal free of biases.

\begin{acknowledgements}
We would like to thank the San Diego Supercomputer Center which allowed us to perform our simulations. A.D. acknowledge support by the Deutsche Forschungsgemeinschaft under Wi~1369/21-1.
\end{acknowledgements}


\bibliography{bibliography,ng-bibs/self,ng-bibs/igm}

\end{document}